\documentclass{article}
\usepackage[T1]{fontenc}
\usepackage[utf8]{inputenc}
\usepackage{ismir} 
\usepackage{amsmath,cite,url}
\usepackage{graphicx}
\usepackage{color}
\usepackage{bbm}
\usepackage{algorithm}
\usepackage{algpseudocode}
\usepackage{amssymb}
\usepackage{multirow}
\usepackage{bm}
\usepackage{tabularx}
\usepackage{array}
\usepackage{ragged2e}

\newcolumntype{P}[1]{>{\RaggedRight\arraybackslash}p{#1}}  
\newcolumntype{X}{>{\RaggedRight\arraybackslash}X}

\title{Identification and Clustering of Unseen Ragas in Indian Art Music}

\oneauthor
  {Parampreet Singh$^{\dag}$, Adwik Gupta, Aakarsh Mishra, Vipul Arora}
  {Indian Institute of technology, Kanpur\\\texttt{params21@iitk.ac.in$^{\dag}$}}

\def\authorname{P. Singh, A. Gupta, A. Mishra and V. Arora}

\begin{document}

\maketitle

\begin{abstract}
Raga classification in Indian Art Music is an open-set problem where unseen classes may appear during testing. However, traditional approaches often treat it as a closed set problem, rejecting the possibility of encountering unseen classes. In this work, we try to tackle this problem by first employing an Uncertainty-based Out-Of-Distribution (OOD) detection, given a set containing known and unknown classes.
Next, for the audio samples identified as OOD, we employ Novel Class Discovery (NCD) approach to cluster them into distinct unseen Raga classes. We achieve this by harnessing information from labelled data and further applying contrastive learning on unlabelled data. 
With thorough analysis, we demonstrate the influence of different components of the loss function on clustering performance and examine how varying openness affects the NCD task in hand. 

\end{abstract}

\section{Introduction}\label{sec:introduction}
Ragas form the core melodic framework of Indian Art Music (IAM), each characterized by a distinct set of notes and improvisational rules that evoke specific emotions or moods\cite{Serra_computational_2017_raga}. Identifying Ragas in audio recordings has various applications, including music recommendation systems, cultural preservation, and music education \cite{Serra_computational_2017_raga}. While traditional methods would rely on handcrafted features and expert knowledge, recent advancements in deep learning have enabled automated Raga identification\cite{phononet_raga,multimodal_HCM_raga,bidkar12north_ensemble_raga,Madhusudhan2024DeepSRGMS_raga,Paramsingh2024explainabledeeplearninganalysis}, where the shortage of labeled datasets remains a significant challenge. Labeling Raga audios in MIR is costly and labor-intensive, requiring domain expertise, while variations in style and recording conditions further complicate annotation.
The problem of Raga identification is inherently an open-set problem, since the number of Ragas is not fixed, and new, unseen classes can emerge during testing, making classification more challenging. 
However, existing approaches have largely treated it as a closed-set problem\cite{phononet_raga,multimodal_HCM_raga,Madhusudhan2024DeepSRGMS_raga,Paramsingh2024explainabledeeplearninganalysis}, limiting their ability to handle novel Raga classes during testing.

This work tackles the challenge of unknown Raga classes through the following approach. First, we perform Out-of-Distribution (OOD) detection by using uncertainty estimates from a model trained only on seen classes, identifying unseen Ragas without prior exposure to them.
Next, we frame this as a Novel Class Discovery (NCD) problem, where the OOD Raga samples are assumed to belong to distinct, previously unseen classes and are clustered in a self-supervised manner.
For NCD, we would generally have target classes $<=$ training classes. So, we define openness of the NCD problem in a similar manner to open-set\cite{open-set_walter_TPAMI} problems as:
\begin{equation}\label{eq:Openness}
\text{O}_{\text{NCD}} = 1 - \sqrt{\frac{2 \times |\text{training classes}|}{2 \times|\text{training classes}| + |\text{test classes}|}}
\end{equation}

For our task, we define two disjoint subsets of Raga classes: a closed-set training set $\mathcal{C}_\text{train}$ consisting of 12 known Raga classes belonging to PIM~\cite{Paramsingh2024explainabledeeplearninganalysis} dataset, and a held-out target set $\mathcal{C}_\text{test}$ comprising novel Raga classes that are entirely unseen during training, belonging to both Saraga (Hindustani)\cite{Saraga} and PIM~\cite{Paramsingh2024explainabledeeplearninganalysis} datasets. 
We analyze our approach on varying levels of openness on both the datasets.

By utilizing this framework, we can effectively tap into the vast amount of freely available, unlabeled Raga recordings from online platforms like YouTube, significantly reducing dependence on manually labeled data. Our approach not only addresses the challenge posed by limited labeled datasets but also enhances the ability of MIR systems to recognize a broader range of Ragas, providing a scalable and adaptive solution for music classification. The codes, metadata can be accessed at:\\ \textit{github.com/ParampreetSingh97/NCD\_ISMIR\_2025.git}

\section{Related Works}
\subsection{OOD Detection}
Uncertainty estimation is a well-established field in machine learning that focuses on evaluating the confidence of model predictions for given test examples.
Various approaches utilize uncertainty for identifying OOD samples. The work\cite{hendrycks2016uncertaintybaseline} proposes using maximum softmax probabilities as uncertainty indicators. Deep ensembles\cite{lakshminarayanan2017deepensamblesimple} combine multiple models to achieve robust uncertainty estimates. Bayesian Neural Networks offer principled uncertainty quantification through posterior distribution approximation. Other methods include techniques which train an auxiliary model to predict confidence scores\cite{corbiere2019_confidnet,corbiere2021confidence_confidnet_2,wimaga_param}. Monte Carlo dropout (MC-dropout)\cite{gal2016mcdropout} applies dropout during inference to simulate Bayesian sampling. In our work, we utilize uncertainty scores from MC-dropout for OOD detection, leveraging our pre-trained model without requiring additional training.

\subsection{Novel Class Discovery (NCD)}
Novel Class Discovery focuses on clustering unknown classes in unlabeled data while utilizing knowledge from labeled data of known classes\cite{object_graphs,ranking_statistics_ICLR,Zhong_2021_CVPR_NCD_NCL,Han2019_ICCV_deep_transfer_clustering,hsu2019_ICLR_multiclass_without_labels}. Unlike semi-supervised learning\cite{semi_sup1,semi-sup-2}, which assumes shared label spaces, or zero-shot learning\cite{zero-shot,zero-shot-2}, which requires human-defined semantic attributes, NCD enables discovery of novel categories without such dependencies. This makes it particularly valuable for music applications, where new classes continuously emerge.

In the image domain, NCD approaches have explored various contrastive learning techniques. Han et al.\cite{Han2019_ICCV_deep_transfer_clustering} introduces a graph-based approach for transferring knowledge from labeled to unlabeled data. Ranking statistics\cite{ranking_statistics_ICLR} have been introduced to construct negative samples for contrastive loss, while Neighborhood Contrastive Learning (NCL)\cite{Zhong_2021_CVPR_NCD_NCL} replaces ranking statistics with cosine similarity and proposes methods for generating hard negatives. 

\subsection{Self-supervised Learning in Music Classification}
Several works in music classification have explored self-supervised learning techniques.
Differentiable ranking\cite{Learn_audio_represent_SPL_2021} techniques on spectrogram patches improve instrument classification and pitch estimation, though this approach is computationally intensive.
\cite{Masataka_goto_self_supervised_singing_voice} utilizes self-supervised contrastive learning for singing voice analysis by applying audio-specific transformations such as time-stretching and pitch-shifting to distinguish vocal timbre and expression. Another study, \cite{s3t_ICASSP_swin_transformer}, integrates the Swin Transformer into a contrastive learning framework for music genre classification, demonstrating strong performance with limited labeled data. Additionally, \cite{ICASSP_2021_unsup_cont_learning} explores the reordering of shuffled spectrogram segments to improve learned audio representations for tasks such as instrument classification and pitch estimation.

In our work, for NCD, we build on Neighborhood Contrastive Learning (NCL)\cite{Zhong_2021_CVPR_NCD_NCL} with tailored modifications in positive/negative pair generation and better transformations for consistency loss for our task. We train a supervised model to learn meaningful representations, and then use these representations to train another model in a self-supervised manner to discover and categorize novel Raga classes in the unlabeled dataset.

\begin{figure}[htbp]
\includegraphics[width=0.9\columnwidth]{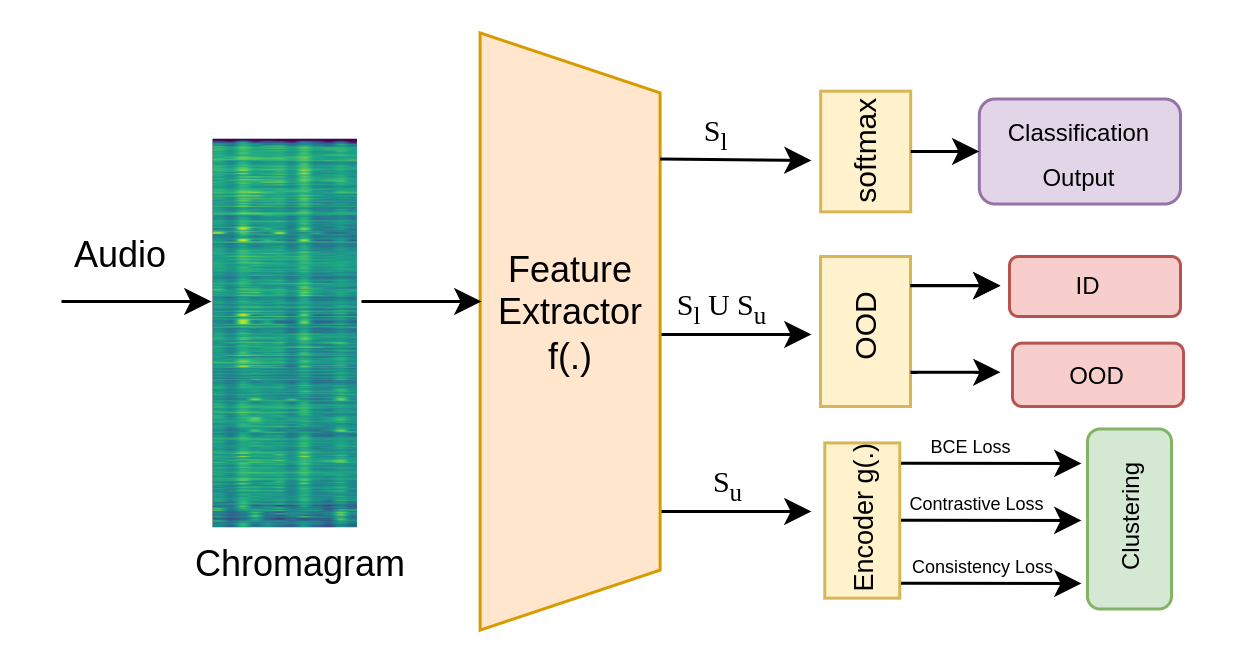}
\caption{Block diagram illustrating the overall system workflow: audio input is first converted to a chromagram and processed by a feature extractor. Extracted features are then used for classification, out-of-distribution (OOD) detection, and subsequent clustering of OOD samples, enabling both in-distribution classification and unsupervised grouping of OOD data.}
\label{fig:blck_dgrm}
\end{figure}

\section{Method}

The overall flow of the whole process is shown in Figure~\ref{fig:blck_dgrm}.
We construct a labeled subset \( \mathcal{S}^l \) containing N number of 30-second audio clips \( \mathbf{x}_i^l \), sourced from the PIM dataset \cite{Paramsingh2024explainabledeeplearninganalysis}, each belonging to one of the $c^t$ predefined Raga classes. We pre-process to remove speech segments, discard audio clips shorter than 30 seconds, and subsequently extract tonic-normalized chromagram features \cite{Paramsingh2024explainabledeeplearninganalysis}, which forms the input to train the Raga classifier $f(\cdot)$. Formally, the labeled subset \( \mathcal{S}^l \) is defined as:  
$\mathcal{S}^l = \left\{ (\mathbf{x}_i^l, c_i^t) \right\}_{i=1}^{N},$
where 
\( c_i^t \in \mathcal{C}^l \) corresponds to its ground-truth Raga label.

Similarly, we define an unlabeled subset of M samples $\mathcal{S}^u = \left\{ \mathbf{x}_i^u \right\}_{i=1}^{M},$
where the corresponding class labels are assumed to be absent. The set of unseen classes \( C^u \) 
is varied in size based on the openness of the problem. 

\subsection{Supervised pre-training}
For classification, we split $S^l$ into training, validation, and test subsets, and train a CNN-LSTM model $f(\cdot)$ in a fully supervised manner using categorical cross-entropy loss. 
Once trained, this CNN-LSTM model serves as a feature extractor by removing the final softmax layer. The resulting feature extractor, denoted as 
$f_{feat}(\cdot)$, generates embeddings $y_i$
for both $S^l$
and $S^u$, which are later used for OOD detection and NCD.

\subsection{OOD Detection}
Monte Carlo (MC) Dropout\cite{gal2016mcdropout} is a technique for estimating epistemic uncertainty in deep learning models. Given a pre-trained CNN-LSTM classifier $f(\cdot)$, we enable dropout at inference time to approximate a Bayesian neural network. The predictive uncertainty is estimated by performing $T$ stochastic forward passes, yielding a set of softmax outputs.
By doing this, we assume that the network's parameters $\mathcal{W}_t$ vary under different dropout masks.
The variance of these predictions quantifies uncertainty values.
Higher variance indicates greater uncertainty, suggesting a higher likelihood of the sample belonging to an OOD class.

\subsection{Novel Class Discovery}
\subsubsection{BCE Loss}
For an input audio clip $x_i^u \in S^u$, let $y_i^u=f_{feat}(x_i^u)$ be the embeddings using the pre-trained feature extractor $f_{feat}(\cdot)$. 
The cosine similarity \( \varepsilon \) between a pair of feature embeddings \( (y_i^u, y_j^u) \) is given by:
\begin{equation}
\varepsilon(y_i^u, y_j^u) = \frac{(y_i^u)^\top y_j^u}{\|y_i^u\| \|y_j^u\|}
\end{equation}
Now, using this, we assign a pairwise pseudo-label \( t_{i,j} \) as:
\begin{equation}
t_{i,j} = \mathbbm{1}{\left[ \varepsilon(y_i^u, y_j^u) \geq \delta \right]},
\end{equation}
where \( \delta \) is a similarity threshold that determines whether the two samples belong to the same latent class. Furthermore, if two audio samples \( x_i^u \) and \( x_j^u \) are formed by splitting from the same audio file, they are assigned \( t_{i,j} = 1 \) as they definitely belong to the same class.  

These pairwise pseudo-labels are used to train a self-attention encoder model \( g(\cdot) \), which incorporates a multi-head self-attention mechanism utilizing scaled dot-product attention, along with layer normalization and feedforward sub-layers. The network consists of multiple such stacked layers, with the input being the embedding $y_i^u$ and its output denoted as $z_i^u=g(y_i^u)$. For BCE loss between the given pair of inputs, we define normalized dot product between the output embeddings, given by $p_{i,j}$:

\begin{equation}\label{eq:pij_bce}
p_{i,j} = \frac{(z_i^u)^\top z_j^u}{\|z_i^u\| \cdot \|z_j^u\|}
\end{equation}
The BCE loss function is defined as:
\begin{equation}\label{eq:bce_loss}
\ell_{bce} = t_{i,j} \log (p_{i,j}) + (1 - t_{i,j}) \log (1 - p_{i,j}).
\end{equation}

\begin{algorithm}
\caption{alg:Novel Raga Clustering}
\label{alg:your_algo2}
\begin{algorithmic}[1]
\Require OOD dataset $S^u$, feature extractor $f_{feat}(\cdot)$
\Require Encoder model $g(\cdot)$, learning rates $\beta, \gamma$, temperature parameter $\tau$
\State Extract chromagram features from all $x_i^u \in S^u$
\State Use $f_{feat}(\cdot)$ to compute embeddings $y_i^u$
\For{each pair $(y_i^u, y_j^u) \in S^u$}
    \State Compute cosine similarity $\varepsilon(y_i^u, y_j^u)$
    \State Assign pseudo-label $t_{i,j}$ based on threshold $\delta$
    \State Compute $p_{i,j}$ using Eq: \ref{eq:pij_bce}
    \State Compute BCE Loss $\ell_{bce}$
\EndFor
\For{each sample $x_i^u$}
    \State Apply time and volume shifts on $x_i^u$ to get $\hat{x}_i^u$
    \State Compute transformed embeddings $\hat{y}_i^u = f(\hat{x}_i^u)$
    \State Compute Consistency Loss $\ell_{mse}$
\EndFor
\For{each sample $x_i^u$}
    \State Select $H$ hardest negative samples $\xi_m$ and positive samples $\phi$
    \State Define contrastive loss $\ell_{cl}$ using positive and negative pairs
\EndFor
\State Compute total loss: $\ell = \ell_{bce} + \beta\ell_{cl} + \gamma\ell_{mse}$
\For{epoch = 1 to $E$}
    \State Train $g(\cdot)$ using total loss $\ell$
\EndFor
\State \textbf{Output:} Trained model $g(\cdot)$
\end{algorithmic}
\end{algorithm}

\subsubsection{Consistency Loss}

To enforce consistency under transformations, we introduce a loss ensuring that an audio sample \( x_i \) and its transformed versions \( \tilde{x}_i \) yield similar outputs. We generate alternate views by time shifting, where for a given audio clip, we create two transformed versions by slightly shifting its start and end times (by 2 seconds) within the original audio, and by volume modification (increase and decrease). We then extract embeddings from the transformed audio \( \tilde{x}_i \), obtaining \( \tilde{z}_i = g(f(\tilde{x}_i)) \), and apply MSE loss as:

\begin{equation}\label{eq:mse_loss}
\ell_{mse} = \frac{1}{C^l} \sum_{i=1}^{C^l} \left( z_i^l - \tilde{z}_i^l \right)^2 + \frac{1}{C^u} \sum_{j=1}^{C^u} \left( z_j^u - \tilde{z}_j^u \right)^2.
\end{equation}

\subsubsection{Contrastive Learning}
To define contrastive loss, we construct positive and negative pairs for our dataset.
For negative pairs, each \( y_i^u \) is compared with all embeddings \( y_n \in S^u \cup S^l \), using cosine similarity $\varepsilon(y_i^u, y_n)$. We then create a list $\zeta_i^u$.

\begin{equation}
\zeta_i^u = \text{list}(\varepsilon(y_i^u, y_n)), \quad \forall \{y_i^u \in S^u\}.
\end{equation}
The similarities are ranked in ascending order, and the H least similar embeddings are selected as hard negatives:
\begin{equation}
\xi_h = \text{argtop}_h (\zeta_i^u), \quad \forall i.
\end{equation}

For positive pairs, similar to BCE loss, all the audio samples \( x_i^u \) and \( x_j^u \) originating from same audio file 
are considered to belong to the same class and hence treated as positive pairs. Their corresponding embeddings \( \hat{z}_i^u \) are stored in the set \( \phi \) which is defined as: 

\[\phi = \{ \hat{z}_i^u \mid z_i^u \text{ shares the same source audio file} \}\]

We now define the contrastive loss \( \ell_{cl} \)~\cite{Zhong_2021_CVPR_NCD_NCL} as:

\begin{equation}\label{eq:contrastive_loss}
\ell_{cl}= -\frac{1}{k} \sum_{\hat{z}_i^u \in \beta} \log \frac{e^{\varepsilon(z_i^u, \hat{z}_i^u) / \tau}}{e^{\varepsilon(z_i^u, \hat{z}_i^u) / \tau} + \sum_{\bar{z} \in \xi_m} e^{\varepsilon(z_i^u, \bar{z}_m^u) / \tau}},
\end{equation}

where \( \tau \) is a temperature parameter that controls the concentration of similarity scores. This loss function optimizes embeddings by bringing each sample closer to its positive counterpart \( \hat{z}_i^u \) while pushing it away from hard negatives \( \bar{z}_m^u \).
Finally, get a unified objective function $\ell$ by combining eq:~\ref{eq:bce_loss},\ref{eq:mse_loss},\ref{eq:contrastive_loss}:

\begin{equation}\label{eq:final_loss}
\ell = \ell_{bce} + \beta \ell_{cl} + \gamma \ell_{mse}.
\end{equation}
Here, \( \beta \) and \( \gamma \) are the scaling hyperparameters, as the magnitude of these losses varies significantly. Proper tuning of these hyperparameters is critical for achieving optimal performance. This combined loss is used to train the self-attention encoder $g(\cdot)$, which learns to differentiate unseen raga classes in a self-supervised manner. This whole training process is explained in Algorithm~\ref{alg:your_algo2}.

\subsection{Clustering Techniques}\label{subsec:clustering_methods}
Given the embeddings \( z_i \) from the encoder model \( g(\cdot) \), 
we experiment with three different approaches for grouping the embeddings to assign predicted labels:\\
(i) Computing a cosine similarity matrix across embedding pairs (\( z_i, z_j \)), and given a threshold $th$, grouping those with similarity \( \varepsilon > th \) into the same cluster.\\
(ii) Applying K-means clustering to group the embeddings into \( K \) clusters.\\
(iii) Reducing the dimensionality of embeddings using UMAP for visualization, followed by K-means clustering on the transformed representations.

\subsection{Evaluation Metrics}\label{subsec:Evaluation Metrics}
We assess the quality of the clusters so formed using both label-independent and label-dependent evaluation metrics.\\
\textbf{(i)} \textbf{Silhouette Score}(SS)~\cite{ROUSSEEUW198753_silhoutte_score} is a label-independent metric, which evaluates how well a data point is situated within its designated cluster in relation to other clusters, without considering the ground truth for those clusters. The score falls between -1 and 1. For well-separated clusters, SS comes out to be 1, and it is -1 for poorly formed clusters. \\
\textbf{(ii)} \textbf{Adjusted Rand Index} (ARI)\cite{JMLR_ARI_rand_2010} is a label-dependent metric, which compares the similarity between predicted clusters and actual ground truth clusters, with an adjustment for random assignments. The score ranges from 0 to 1, where 1 represents perfect alignment with the ground truth.\\
\textbf{(iii)} \textbf{Mutual Information} (MI)\cite{NMI} measures the amount of information shared between the true clusters ($c^t$) and predicted clusters ($c^p$). It captures how much knowing the predicted cluster assignment reduces uncertainty about the true cluster assignment. The range of MI is not bounded, with higher values indicating that the predicting clustering is more aligned with the actual class structure.\\
\textbf{(iv)} \textbf{Clustering Accuracy} (ACC) evaluates how well the predicted clusters align with the true labels. For each ground truth cluster \( c^t \), we identify the predicted cluster \( c^p \) that has the highest overlap with \( c^t \). The subset of embeddings that belong to both \( c^t \) and \( c^p \) is represented as:
\[
c^{pt} = \{ z_i \mid z_i \in c^p \text{ and } z_i \in c^t \}.
\]
Then, ACC for a given true cluster \( c^t \) is then computed as:
\[
\text{ACC}(c^t) = \frac{|c^{pt}|}{|c^t|} \times 100.
\]
Misclassified points are those that do not belong to any matched cluster. Furthermore, if a predicted cluster \( c^p \) is mapped to multiple true clusters \( c^t \), the clustering is considered invalid, and accuracy, along with other performance metrics, is not calculated.

\section{Experimental Results}

The labeled dataset 
$S^l$ consists of 141 audio files sourced from PIM~\cite{Paramsingh2024explainabledeeplearninganalysis} dataset, segmented into 5,734 audio samples, with a total duration of approximately 47.78 hours. A CNN-LSTM model $f(\cdot)$ is trained in a supervised manner on this dataset for multi-class classification across 12 Raga classes, achieving an F1-score of 0.89 through cross-validation. This trained model serves as a feature extractor for downstream tasks, where representations for OOD detection and NCD are obtained by extracting features from different depths of the network.
We construct another set $S^u$
for which the Raga labels are discarded, treating it as unlabeled data. 
We conduct a range of OOD and NCD experiments using both the PIM~\cite{Paramsingh2024explainabledeeplearninganalysis} and Saraga~\cite{Saraga} (Hindustani) datasets at different stages, as summarized in Table~\ref{tab:overall_experiments}.

\begin{table}[h]
\centering
\renewcommand{\arraystretch}{1.2}
\begin{tabularx}{\linewidth}{|P{1.9cm}|P{0.95cm}|X|}
    \hline
    \textbf{Experiment} & \textbf{Dataset} & \textbf{Description} \\
    \hline
    OOD detection & PIM/ Saraga & Carry out OOD Detection for both datasets separately using $f(\cdot)$; results in Table~\ref{tab:ood_accuracy} \\
    \hline
    Feature ablation & PIM& Compare Chromagram vs Melody\cite{melody_kavya} vs MERT\cite{li2024mert} features; results in Table~\ref{tab:labeled_clusters_MERT} \\
    \hline
    Loss component ablation & PIM & Test $\ell_{bce}/\ell_{cl}/\ell_{mse}$ contributions; results in Table~\ref{tab:loss_comparison} \\
    \hline
    Clustering comparison & PIM/ Saraga & Evaluate Cosine-sim vs K-Means vs UMAP+K-means; results in Table~\ref{tab:comparison_pim_saraga} \\
    \hline
    Openness study & PIM & Analyze performance at openness = 0.09 \& 0.18; results in Table~\ref{tab:comparison_openness} \\
    \hline
\end{tabularx}
\caption{Summary of all experimental setups, datasets, and their corresponding result locations in the paper.}
\label{tab:overall_experiments}
\end{table}
\begin{table}[h]
    \centering
    \begin{tabular}{|c|c|c|}
        \hline
        \textbf{Metric/Dataset} & \textbf{Saraga} & \textbf{PIM} \\
        \hline
        OOD Accuracy & 85.6\% & 80.87\% \\
        \hline
    \end{tabular}
    \caption{Comparison of OOD detection Accuracy for Saraga and PIM datasets}
    \label{tab:ood_accuracy}
\end{table}

\subsection{OOD}
For OOD detection, we select test files from five unseen classes in the PIM and Saraga datasets, prioritizing those with higher representation. From PIM, we use 41 audio files, resulting in 2,435 audio clips (20.29 hours), belonging to 5 Raga classes: Bageshri, Bhopali, Jog-Kauns, Mishra-Khamaj, and Puriya-Kalyan. From Saraga, 14 audio files, yielding 1,136 audio clips (9.46 hours) belonging to 5 Raga classes: Bhopali, Bhimpalasi, Marwa, Shree, Todi. An equal number of files from 
$S^l$ (only from PIM dataset) is included for comparison. The $f(\cdot)$ model is trained with MC-dropout, with T=50 forward passes for each $x_i$, and a variance-based threshold is applied to classify samples as OOD or in-distribution. Results, presented in Table \ref{tab:ood_accuracy}, demonstrate OOD detection performance. 
The model performs better on Saraga for OOD detection, the reason being since the $f(\cdot)$ model is trained on the PIM dataset, the OOD recordings from PIM may share acoustic similarities with the training data, making OOD detection more challenging. In contrast, the Saraga dataset, recorded in different acoustic environments, serves as a more distinct and thus easier target for OOD detection.

\subsection{Feature Ablation} For $S^l$, we extract embeddings using the pre-trained MERT model \cite{li2024mert}, melody-based embeddings from \cite{melody_kavya}, and the feature extractor $f_{feat}(\cdot)$, which extracts embeddings from the penultimate layer of CNN-LSTM classifier $f(\cdot)$. These embeddings are then clustered using cosine similarity, as described in Section~\ref{subsec:clustering_methods}.

The clustering outcomes for $S^l$ are summarized in Table~\ref{tab:labeled_clusters_MERT}. The results indicate that embeddings from both MERT and melody-based models yield subpar performance, even when evaluated with label-independent metrics. In contrast, 
$f_{feat}(\cdot)$ provides significantly better clustering results. So, we adopt $f_{feat}(\cdot)$ as the feature extractor for the remainder of our study.

\begin{table}[htbp]
\begin{center}
\begin{tabular}{|c|c|c|c|}
\hline
\textbf{Metric} & \textbf{MERT} & \textbf{Melody} & $\bm{f_{feat}(\cdot)}$ \\
\hline
\textbf{SS} & 0.13 & -0.01 & \textbf{0.54} \\
\hline
\textbf{ARI} & 0.00 & 0.08 & \textbf{0.83} \\
\hline
\textbf{MI} & 0.02 & 0.22 & \textbf{1.99} \\
\hline
\textbf{ACC} & 11.15 & 25.04 & \textbf{90.05} \\
\hline
\end{tabular}
\caption{Comparison of MERT, Melody extraction tool (Mel), and $\mathbf{f_{feat}(\cdot)}$ for clustering using k-means on $S^l$}
\label{tab:labeled_clusters_MERT}
\end{center}
\end{table}

\begin{table*}[htbp]
\centering
\begin{tabular}{|c|c|c|c|c|c|c|}
\hline
\textbf{Dataset} & \multicolumn{2}{|c|}{\textbf{Clustering Methods}} & \textbf{SS} & \textbf{ARI} & \textbf{MI} & \textbf{ACC (\%)} \\
\hline
\multirow{6}{*}{\textbf{PIM}} 
& \multirow{2}{*}{Cosine Similarity} & Baseline & 0.22 & 0.50 & 0.87 & 58.96 \\
&  & Proposed &\textbf{ 0.75} & \textbf{0.58} & \textbf{1.17} & \textbf{72.10} \\
\cline{2-7}
& \multirow{2}{*}{K-Means} & Baseline & 0.36 & 0.48 & 0.79 & 70.75 \\
&  & Proposed & \textbf{0.85} & \textbf{0.64 }& \textbf{0.94 }& \textbf{79.34} \\
\cline{2-7}
& \multirow{2}{*}{UMAP} & Baseline & 0.63 & 0.53 & 0.83 & 71.48 \\
&  & Proposed & \textbf{0.79 }& \textbf{0.60} & \textbf{0.85} & \textbf{72.99} \\
\hline
\multirow{6}{*}{\textbf{Saraga}} 
& \multirow{2}{*}{Cosine Similarity} & Baseline & 0.15 & 0.30 & 0.65 & 53.61 \\
&  & Proposed & \textbf{0.71} & \textbf{0.43} & \textbf{0.86} & \textbf{78.37} \\
\cline{2-7}
& \multirow{2}{*}{K-Means} & Baseline & 0.40 & 0.41 & 0.79 & 75.44 \\
&  & Proposed & \textbf{0.82} & \textbf{0.44} & \textbf{0.82} & \textbf{81.04} \\
\cline{2-7}
& \multirow{2}{*}{UMAP} & Baseline & 0.60 & 0.44 & 0.81 & 73.85 \\
&  & Proposed & \textbf{0.66 }& \textbf{0.47 }& \textbf{0.85} & \textbf{78.88 }\\
\hline
\end{tabular}
\caption{Performance comparison of clustering methods on PIM and Saraga Datasets}
\label{tab:comparison_pim_saraga}
\end{table*}

\begin{table}[htbp]
\begin{center}
\begin{tabular}{|c|c|c|c|c|}
\hline
{\textbf{Metric}} & \textbf{$\boldsymbol{\ell_{cl}}$} & \textbf{$\boldsymbol{\ell_{bce}}$} & \textbf{$\boldsymbol{\ell_{cl+bce}}$} & \textbf{$\boldsymbol{\ell}$} \\
\hline
\textbf{SS}  &0.39  &0.59  &0.62  &\textbf{0.85}  \\
\hline
\textbf{ARI} &0.52  &0.55  &0.59  &\textbf{0.64}  \\
\hline
\textbf{MI} &0.76  &0.84  &0.87  &\textbf{0.94}  \\
\hline
\textbf{ACC (\%)} &70.16  &75.43  &76.04  &\textbf{79.34}  \\
\hline
\end{tabular}
\caption{Comparison of clustering metrics for training $g(\cdot)$ using $\ell_{cl}$, $\ell_{bce}$, $\ell_{cl+bce}$, and $l$, after clustering $z_i^u$ using K-means clustering}
\label{tab:loss_comparison}
\end{center}
\end{table}

\subsection{NCD}

\subsubsection{Comparison with baseline}
For the baseline, clustering is performed directly on the embeddings $y_i$ using the three clustering methods described in Section~\ref{subsec:clustering_methods}. In our proposed approach, we train the encoder model $g(\cdot)$ using the combined loss $\ell$ (eq:~\ref{eq:final_loss}) on both the PIM and Saraga datasets. The resulting clustering performance for both baseline and proposed methods is presented in Table\ref{tab:comparison_pim_saraga}.
As expected, the baseline results for \( S^u \) are significantly worse than those for the labeled dataset. This outcome is anticipated since the feature extractor \( f_{feat}(\cdot) \) is not trained on \( S^u \), and \( S^u \) and \( S^l \) contain disjoint Raga classes. Consequently, clustering performance is poor for both label-dependent and label-independent clustering metrics under the baseline.

\begin{figure}[htbp]
\includegraphics[width=0.9\columnwidth]{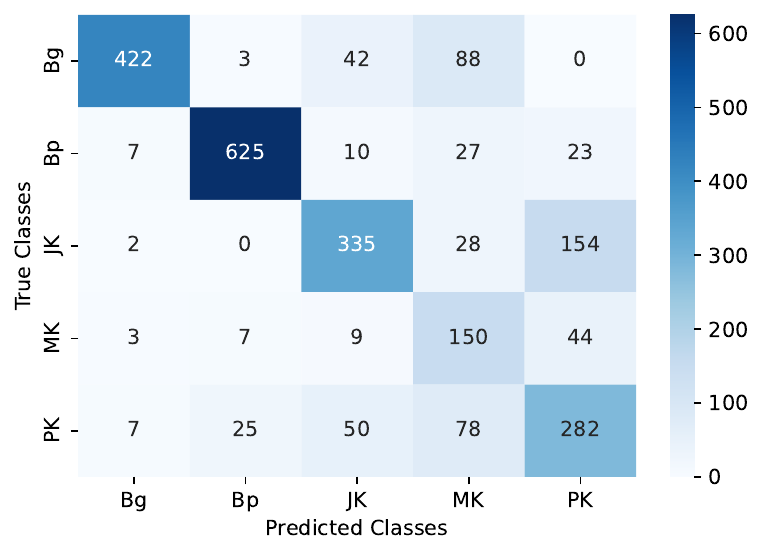}
\caption{Confusion matrix for $S^u$ showing classification performance on the PIM dataset for five Ragas: Bhopali (Bp), Bageshri (Bg), Jog-Kouns (JK), Mishra-Khamaj (MK), and Puriya-Kalyan (PK).}
\label{fig:confusion_matrix}
\end{figure}

Fig. \ref{fig:confusion_matrix} shows the confusion matrix for classification of 5 unknown Raga classes: Bhopali, Bageshri, Jog-Kouns, Mishra-Khamaj, and Puriya-Kalyan out of PIM dataset. We compute f1-scores based on the confusion matrix, and observe that the model performs well for Bageshri (F1: 0.85) and Bhopali (F1: 0.92), which are more distinct and straightforward Ragas. However, it struggles with Mishra-Khamaj (F1: 0.51), Jog-Kouns (F1: 0.69), and Puriya-Kalyan (F1: 0.60). These Ragas, being Mishra (mixed) Ragas, inherently share musical similarities with more than one Ragas in their structure itself, making them more challenging to distinguish and often leading to confusion for the model. This highlights the intrinsic complexity of Mishra Ragas and emphasizes the need for more refined approaches to accurately classify such Ragas.

For the Saraga dataset, trained on Raga Bhopali, Bhimpilasi, Marwa, Todi, and Shree in their set $S^u$, the confusion matrix (not shown) reveals significant overlap between Raag Shree and Marwa. This can be attributed to their structural similarities as they both belong to the Marwa \textit{thaat\footnote{A \textit{thaat} is a parent scale in Hindustani Music, that defines the set of notes used in ragas. If two ragas belong to the same thaat, they are likely to share similar notes, making them more acoustically similar.}}, share common notes with one exception, omit \textit{Pancham\footnote{The fifth note in the scale; when omitted in the Aaroh of ragas from the same Thaat, it further reduces their melodic distinctiveness.}} note in Ascent (\textit{Aaroh}), and are sung at the same time of the day. We also find that the audio recordings for these 2 Ragas feature the same singers in the dataset, and also from the same concert, leading to shared tonal and acoustic characteristics, which may have caused them to cluster closely and, hence, poorer clustering performance compared to the PIM dataset. Another thing is that in Saraga dataset, the representation of each Raga class is limited to max 3 audio files, wherever in PIM, we have at least 7 audio files for each of the unlabeled classes.

\subsubsection{Loss component Ablation}
To understand the individual contributions of different components in our final loss function $\ell$, we train the encoder model $g(\cdot)$ separately using each component—Binary Cross-Entropy (BCE) loss ($\ell_{bce}$), Contrastive loss ($\ell_{cl}$), their sum ($\ell_{cl+bce}$), and the full combined loss $\ell$ (Eq.~\ref{eq:final_loss}). 
For this comparison, we apply K-means clustering on the resulting embeddings using only the PIM dataset. The clustering performance for each setup is summarized in Table\ref{tab:loss_comparison}.

We observe that $\ell_{cl}$ forms poor clusters, as evident from the plot (not shown), where we see all the samples separated like they are plotted along the boundary of a circle. It has been explained by \cite{wang2021understanding_contrastive} also that contrastive Learning (CL) pushes dissimilar samples apart without preserving semantic structure, sometimes grouping unrelated samples while separating similar ones, which is evident here also. 
BCE performs better by focusing on confidently similar pairs and ignoring uncertain ones. $\ell_{cl+bce}$ combines the strengths of both, further improving clustering. Adding MSE enhances semantic consistency, making $\ell$ the most effective, outperforming all three across all metrics.

\subsubsection{Openness Study}
We analyze the impact of openness on clustering performance. As defined in Section~\ref{sec:introduction}, openness is determined by the number of labeled classes $|C^l|$ and the number of unseen classes $|C^u|$. 
In our case, $|C^l|$ is fixed to 12, but we now experiment with values 5 and 12 for $|C^u|$, resulting in openness values of 0.09 and 0.18, respectively for PIM dataset. 
A higher openness value corresponds to a more challenging problem, as is observed in Table~\ref{tab:comparison_openness}. We observe a significant drop in performance, particularly in ACC, suggesting that some classes are being clustered poorly or even randomly, despite a relatively good SS score. This may be due to reduced representation for certain classes as the number of samples per class decreases. Increasing the sample size could potentially improve clustering performance.

\begin{table}[htbp]
\begin{center}
\begin{tabular}{|c|c|c|}
\hline
\textbf{Metric} & \textbf{$\boldsymbol{O_{NCD}=0.09}$} & \textbf{$\boldsymbol{O_{NCD}=0.18}$} \\
\hline
\textbf{SS}  &0.85  &0.50  \\
\hline
\textbf{ARI} &0.64  &0.44  \\
\hline
\textbf{MI} &0.94  &0.83  \\
\hline
\textbf{ACC (\%)} &79.34  &55.68  \\
\hline
\end{tabular}
\caption{Clustering Comparison for Different Levels of Openness Eq:~\ref{eq:Openness} ($O_{NCD}$ )}
\label{tab:comparison_openness}
\end{center}
\end{table}


Our results show that the clusters obtained using the proposed method can approach or match the clustering quality of supervised methods. This is especially valuable in fields like MIR, particularly for the Raga Identification task, where labeled data is limited. 
Our method opens avenues for utilizing vast amounts of unlabeled recordings available online or from various sources, enabling scalable Raga Identification and expanding the repository of Raga datasets without relying heavily on manual labeling.

\section{Conclusion and Future Scope}

In this study, we propose a novel approach for identifying and clustering unseen Raga classes in Indian Art Music. We first use Uncertainty Estimation for Out-of-Distribution (OOD) detection on both the Saraga and PIM datasets, effectively distinguishing unknown Ragas from known ones. Then, we apply a contrastive learning-based Novel Class Discovery (NCD) method in a self-supervised setting to cluster the OOD Ragas into distinct clusters. Our approach demonstrates strong cross-dataset generalization, as features extracted from PIM were successfully used to train and cluster for Saraga. Additionally, we analyze the impact of varying openness values, showing that higher openness yields poorer clustering performance, highlighting the need for further improvements.

Future work can focus on improving the handling of Mishra Ragas, reducing their confusion with their parent Ragas. Expanding Raga Identification datasets will prove beneficial in such applications as we observe a decline with increased openness, partly due to limited class representation. Exploring multimodal techniques or hierarchical representation learning could improve adaptability, making the model more robust to higher openness levels. Additionally, solving NCD as a General Class Discovery (GCD) problem, where the model learns from both labeled and unlabeled sets simultaneously, rather than treating them separately, could be a good future direction. 

\bibliography{ISMIRtemplate}

\end{document}